%% file: arxiv.tex
\documentclass{article} %
\usepackage{iclr2027_conference,times}

\input{math_commands.tex}

\usepackage{hyperref}
\usepackage{url}

\usepackage{amsmath,amssymb,booktabs,graphicx,microtype,xcolor,enumitem,multirow}
\usepackage{flafter}
\usepackage{hyperref,url}
\definecolor{dynblue}{HTML}{DCEBFA}
\definecolor{physorange}{HTML}{FBE4CB}
\definecolor{datagreen}{HTML}{DDEFD8}
\definecolor{learnpurple}{HTML}{E9DDF2}
\definecolor{projred}{HTML}{F5D4D4}

\graphicspath{{assets/}{./}}
\newcommand{\co}{CO$_2$}

\title{AEGIS: Differentiable Mars Climate Model with Neural Closures}

\iclrfinalcopy

\author{Sameera S Kashyap, Victor Cruz, Angel Yepez  \& Razvan Marinescu \\
Department of Computer Science\\
UC Santa Cruz\\
Santa Cruz, CA 95064, USA \\
\texttt{\{ssudars1, vcruz17, ayepezal, ramarine\}@ucsc.edu} \\
}

\begin{document}
\maketitle
\begin{abstract}                                                      
General circulation models (GCMs) are the primary tool for simulating planetary atmospheres. They play a vital role in understanding Mars's atmosphere, as forecasting its unique weather is mission-critical for operations such as entry, descent, and landing. Mars poses unusual challenges for these models, as observations are sparse compared to Earth.  In addition, a thin \co{} atmosphere alongside a radiatively active dust cycle creates a volatile atmosphere with large diurnal temperature swings and no true terrestrial analog for validation. Existing Mars GCMs, including the LMD PCM, the NASA Ames Mars GCM, and PlanetWRF, are mature and physically detailed but are implemented in legacy Fortran with finite-difference or finite-volume solvers, and they do not expose gradients for calibration or machine learning. Here we present AEGIS, a modular differentiable Mars climate model that couples Mars's unique atmospheric physics to the Dinosaur dynamical core, with interfaces for neural closures. We showcase stable ten-Mars-year simulations that reproduce the seasonal \co{} cycle while conserving the total \co{} inventory, capture realistic large-scale surface-temperature structure, and produce surface pressure that follows Mars Orbiter Laser Altimeter (MOLA) topography. Gradients through coupled trajectories agree with finite differences and support physical calibration and neural training. We compare with conventional GCMs, highlighting the framework's differentiability.
\end{abstract}

\section{Introduction}
General circulation models (GCMs) connect the governing equations of atmospheric motion to the physical processes that shape planetary climate. Established Mars models include the LMD Mars GCM, the NASA Ames Mars GCM, and PlanetWRF \citep{forget1999,haberle1993,richardson2007}. On Mars, radiative heating by dust and \co{}, exchange with a thermally active surface, and seasonal condensation onto polar caps jointly shape circulation, pressure, and frost. A useful model must represent these interactions while allowing individual assumptions to be tested \citep{forget2010co2cycle}.

Traditional GCMs contain uncertain parameterization coefficients whose
coupled effects can complicate tuning and permit compensating errors
\citep{forget1999}. Parameter sweeps and ensemble calibration are often computationally expensive \citep{cleary2020augmented}. In addition, the forward-model interface does not generally expose gradients, and maintaining separate adjoint implementations incurs substantial development effort \citep{giering1998}. Introducing a learned component can also require substantial changes to the coupling between numerical and physical routines. These challenges motivate models in which physical simulation and gradient optimization share the same implementation \citep{davenport2026}.

Differentiable programming bridges this gap. When a simulation is
implemented as differentiable numerical operations, automatic differentiation applies the chain rule through its computational graph to obtain derivatives of outputs with respect to inputs and parameters or learned weights \citep{baydin2018autodiff}. This supports calibration, sensitivity analysis, inverse problems, and training over multiple time steps. In particular, a neural parameterization can be optimized for its effect on a coupled trajectory, accounting for feedback between the learned tendency and the resolved dynamics \citep{rasp2020, kashyap2024modelingchaoticlorenzode}. NeuralGCM demonstrates the potential of this approach for Earth's weather and climate by coupling learned parameterizations to a differentiable atmospheric dynamical core \citep{kochkov2024}. Its Dinosaur core solves the hydrostatic primitive equations using spectral horizontal coordinates and a sigma vertical coordinate. The JAX Circulation Model (JCM) uses the same core with an explicit suite of intermediate-complexity physical parameterizations, providing a setting for process-based studies and hybrid modeling for Earth's climate and atmosphere \citep{gelbrecht2023differentiable, davenport2026}. In a different physical domain, JAX MD illustrates how reusable numerical components and neural models can be composed within a common differentiable simulation environment \citep{schoenholz2020}. AEGIS shares the Dinosaur backbone with JCM and NeuralGCM, but addresses a different physical regime and adds coupled state variables and operators needed for Mars.

AEGIS couples Mars radiation and surface physics to Dinosaur's differentiable dynamical core, using Mars radius, rotation, and thermodynamic scales at the dynamical interface. On Mars, condensation of \co{}, the main atmospheric constituent, transfers a substantial fraction of atmospheric mass to surface frost and changes surface pressure \citep{forget2010co2cycle}. AEGIS feeds this exchange into prognostic log surface pressure and evolves frost and regolith temperatures alongside the atmosphere. Automatic differentiation passes through both the coupled evolution and the per-step atmosphere--frost inventory projection, so parameter sensitivities include the pressure response and its reservoir constraint.

We make three contributions:
\begin{enumerate}[leftmargin=*,nosep]
\item We couple Dinosaur's atmospheric dynamics to Mars surface, frost, and regolith states, with \co{} mass exchange entering prognostic pressure. Differentiation through the atmosphere--frost inventory projection preserves the total-inventory constraint in parameter sensitivities.
\item We provide Mars radiation, surface, and transport operators with interfaces for bounded neural heating and learned radiative fluxes. The radiation interface computes heating from flux convergence, preserving the discrete column radiation budget for any network weights.
\item We test physical and neural gradients against finite differences, evaluate Mars fields against reference products, and report long integrations and pilot experiments in physical calibration and neural training (Section~\ref{sec:results}).
\end{enumerate}

\section{Methods}
AEGIS couples a modular Mars physics suite to the Dinosaur spectral dynamical core, carrying the atmosphere, surface, \co{} frost, and regolith in a single differentiable state (Section~\ref{sec:framework}). Automatic differentiation through this coupled time step provides gradients for physical calibration and for training learned components that enter through a common neural interface (Section~\ref{sec:differentiation}).

 \subsection{Atmospheric configurations and the dynamical framework}\label{sec:framework}

AEGIS separates the description of a planetary atmosphere from the numerical machinery used to evolve it. A body configuration specifies radius $a$, gravity $g$, rotation rate $\Omega$, specific gas constant $R$, heat capacity $c_p$, and reference temperature and pressure. We use Mars values for the planetary constants \citep{nasaMarsFactSheet}. The dynamical equations consume these values through a common interface. Length and time are scaled by $a$ and $(2\Omega)^{-1}$, respectively, so the nondimensional radius and rotation rate follow Dinosaur's conventions for each body, removing dependence on Earth-specific default scales while retaining the existing spectral operators.

The framework supports a hierarchy of atmospheric configurations. Dynamical configurations isolate resolved transport; physically forced configurations add radiation, surface exchange, mixing, and volatile reservoirs; hybrid configurations add a learned tendency or substitute a learned component at a selected physical interface. Complexity is therefore introduced through explicit components: a body definition establishes the mechanical and thermodynamic constants, while its radiation, condensables, and boundary conditions establish the atmospheric regime. The physically developed configuration evaluated here is Mars; extending to another atmosphere involves supplying the relevant process operators as well as its constants \citep{hu2025stablemachinelearningparameterizationsubgrid}.

Dinosaur advances the hydrostatic primitive equations on a sphere \citep{kochkov2024,davenport2026}. The atmospheric state contains spectral vorticity and divergence, temperature relative to a reference profile, log surface pressure, optional transported tracers, and simulation time, on a vertical coordinate $\sigma=p/p_s$ that follows changes in atmospheric mass and the terrain-dependent lower boundary \citep{Richardson2018OnTR}. 
We retain Dinosaur's spherical-harmonic discretization for the horizontal dynamics. The main experiments use triangular truncation T21, which retains spherical harmonics through degree 21, and 12 equally spaced sigma layers. A $64\times32$ longitude--latitude transform grid is used to evaluate nonlinear terms and column physics at geographic locations. Spectral transform models using sigma coordinates have a long history in Mars atmospheric modeling \citep{wilson1997stationarywaves}.

The coupled state $\vx$ contains the atmospheric variables, surface temperature, pressure-equivalent \co{} frost, and a vertical regolith temperature profile for each geographic column. Its rate of change combines resolved atmospheric motion, physical processes, and an optional learned tendency. We write the resolved dynamics as $\mathcal{D}_{\mathrm{imp}}+\mathcal{D}_{\mathrm{exp}}$, with the subscripts identifying the terms treated implicitly and explicitly by the time integrator. The operator $\mathcal{P}_{\boldsymbol{\alpha}}$ gives physical tendencies controlled by parameters $\boldsymbol{\alpha}$, and $\mathcal{N}_{\boldsymbol{\theta}}$ gives the optional neural tendency with network parameters $\boldsymbol{\theta}$. The coupled equations are
\begin{equation}
 \frac{d\vx}{dt}=\mathcal{D}_{\mathrm{imp}}(\vx)
 +\mathcal{D}_{\mathrm{exp}}(\vx)
 +\mathcal{P}_{\boldsymbol{\alpha}}(\vx,t)
 +\mathcal{N}_{\boldsymbol{\theta}}(\vx,t),
 \label{eq:coupled}
\end{equation}
A neural replacement for a physical component is composed inside $\mathcal{P}$ instead of being added a second time. Time integration uses the implicit--explicit SIL3 Runge--Kutta scheme \citep{whitaker2013}, with linear gravity-wave terms treated implicitly and nonlinear dynamics and assembled physical tendencies on the explicit side. Planetary unit conversion, the augmented state, and the atmosphere--frost correction are implemented in AEGIS's surrounding operators.

To control the growth of small-scale numerical noise, we apply eighth-order hyperdiffusion after each time step. This smooths wind and temperature variations most strongly at the shortest resolved scales, helping stabilize the integration while preserving the global-mean mode. A damping timescale of 0.1 sol sets the strength at the shortest resolved scale. The filter, Courant diagnostic, and experiment-specific numerical settings are detailed in Appendix~\ref{app:numerics}.

\begin{figure}[t]
\centering
\includegraphics[width=.65\linewidth]{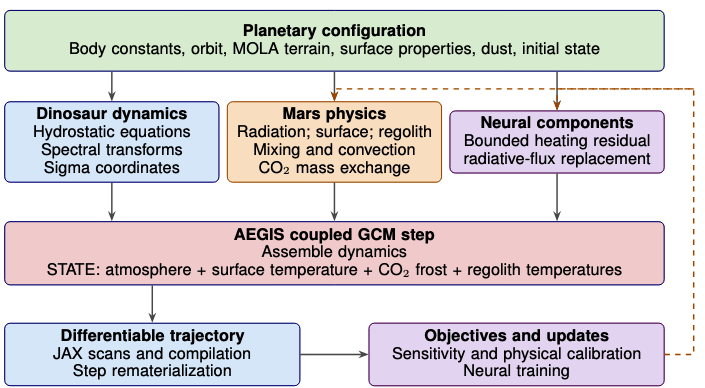}
\caption{AEGIS architecture. Planetary constants and boundary data configure a shared GCM. Dinosaur supplies resolved dynamics; Mars physical operators supply column tendencies and surface evolution; neural components enter as bounded residuals or component replacements. The coupled time step advances atmosphere and surface reservoirs. Trajectory objectives connect the simulated fields to physical-parameter calibration and neural training through automatic differentiation. Dashed arrows denote this gradient-based update path.}
\label{fig:architecture}
\end{figure}

The Mars operators in $\mathcal{P}_{\boldsymbol{\alpha}}$ span \co{} and dust radiation, surface and regolith energy exchange, boundary-layer mixing and dry convection, and a \co{} condensation cycle \citep{lacis1991,forget1999,haberle2019ames,petrosyan2011pbl}. Surface condensation and sublimation transfer mass between the atmosphere and frost. Let $p_s$ denote local surface pressure and $I=g\,m_{\mathrm{ice}}$ denote surface frost mass per unit area expressed in pressure units, where $g$ is gravitational acceleration. Condensation increases $I$ and lowers $p_s$; sublimation reverses this exchange. The resulting contribution to prognostic log surface pressure is
\begin{equation}
 \left.\frac{d\log p_s}{dt}\right|_{\mathrm{phase}}
 =-\frac{1}{p_s}\frac{dI}{dt},
 \label{eq:mass}
\end{equation}

Here, $dI/dt>0$ denotes net condensation from the atmosphere onto the surface, which increases the frost inventory and decreases $p_s$. Conversely, $dI/dt<0$ denotes net sublimation of surface \co{} frost into the atmosphere, which decreases the frost inventory and increases $p_s$. The factor $1/p_s$ converts the surface-pressure tendency to a log-pressure tendency. After each complete time step, a projection repairs negative frost and rescales atmospheric pressure to restore the pre-step atmosphere-plus-frost inventory. This corrects numerical inventory error while allowing physical exchange between the reservoirs. Because \co{} is both a radiative gas and a substantial mobile mass reservoir, carrying its phase change through the prognostic pressure state, alongside surface and subsurface temperatures \citep{kellyseaonalpolar}, makes this coupling part of the simulated system and its derivatives. The radiation, surface, boundary-layer, and \co{}-inventory operators are detailed in Appendix~\ref{app:physics}.

\subsection{Differentiation and neural components}\label{sec:differentiation}

Let $\Phi_{\boldsymbol{\alpha},\boldsymbol{\theta}}$ advance the coupled state by one time step, including reservoir corrections. Repeated updates $\vx_{n+1}=\Phi_{\boldsymbol{\alpha},\boldsymbol{\theta}}(\vx_n)$ produce a trajectory. The operator $\mathcal{H}$ extracts the fields compared with target data $\vy_n$ at step $n$. A trajectory loss $\mathcal{L}(\boldsymbol{\alpha},\boldsymbol{\theta})$ measures their mismatch over steps $1,\ldots,N$. JAX differentiates this discrete numerical trajectory using automatic differentiation. We use forward-mode differentiation for the three physical controls and reverse-mode differentiation for neural weights; forward mode efficiently evaluates derivatives with respect to a small number of controls, whereas reverse mode is appropriate for a scalar loss with many parameters \citep{baydin2018autodiff,jax2024autodiff}. Reverse-mode training through coupled rollouts, as in the heating experiment below, uses fixed-length scans with step rematerialization, which trades additional computation for reduced reverse-mode memory use by recomputing selected forward intermediates during the backward pass \citep{jax2024checkpoint}.

Learned components either add a bounded atmospheric heating tendency or replace the physical radiation operator within the coupled time step. Each network is initialized from a random seed, and input normalization is fitted on training data only. For T21 horizontal truncation (spherical harmonics through degree 21) and 12 vertical sigma layers, denoted T21/L12, the heating network uses 21 column features: the temperature profile, surface temperature and pressure, lowest-layer winds, geographic coordinates, and insolation. Its heating tendency is
\begin{equation}
 \dot T_{\mathrm{NN}}=h_{\max}\tanh f_{\boldsymbol{\theta}}(\vz),
 \label{eq:neural}
\end{equation}
Here, $f_{\boldsymbol{\theta}}$ is the network, $\vz$ contains the normalized column features, and $h_{\max}$ bounds the heating magnitude. The heating is converted to the dynamical core's tendency units and added to the physical tendencies. Surface reservoirs remain under physical control. A zero-initialized output layer reproduces the physical model at the start of training.

The learned radiation operator predicts interface fluxes from column temperature and pressure profiles, dust, sunlight, albedo, and emissivity. Shortwave fluxes are bounded by incoming sunlight, longwave fluxes are nonnegative, and the top and surface boundary conditions are imposed exactly. Heating follows from flux convergence, closing the discrete column radiation budget for any network weights.

Both interfaces support custom networks and check vertical-grid and radiation-configuration compatibility before use. Their parameters can be trained on local targets or through coupled rollouts using reverse-mode differentiation. Validated checkpoints store the learned parameters, normalization, and configuration metadata. Figure~\ref{fig:architecture} shows the overall AEGIS framework, from planetary constants and boundary data to dynamics, physical operators, and neural components. The coupled time step advances the atmosphere and surface reservoirs, with neural components entering as bounded tendencies or physical-operator replacements. Trajectory losses connect the simulated fields to parameter calibration and neural training through the gradient paths shown by dashed arrows. Training protocols for the reported experiments are given in Appendix~\ref{app:protocol}.

\section{Results}\label{sec:results}

\subsection{Gradient checks and their use in calibration}\label{sec:gradient-checks}

\begin{table}[t]
\centering
\small
\caption{Derivative checks through coupled AEGIS trajectories. All
comparisons satisfy
\(\lvert g_{\mathrm{AD}}-g_{\mathrm{FD}}\rvert
\leq10^{-8}+10^{-4}\lvert g_{\mathrm{FD}}\rvert\).}
\label{tab:derivative_checks}
\begin{tabular}{llccc}
\toprule
Category & Objective & Horizon & Checks & Passed \\
\midrule
Physical sensitivity &
Final mean atmospheric temperature &
1--148 steps &
54 &
54/54 \\

Physical calibration &
Scalar fitting objective &
25 steps &
9 &
9/9 \\

Neural fitting &
Multi-field trajectory objective &
25 steps &
3 &
3/3 \\

Multi-diagnostic sensitivity &
Mean surface $T$, $p_s$, and \co{} frost &
74--1480 steps &
9 &
9/9 \\
\bottomrule
\label{diff-table}
\end{tabular}
\end{table}

Gradients quantify how a model output or fitting loss responds to a change in a physical parameter or neural weight. They provide update directions for calibration and training while accounting for feedbacks along the coupled trajectory. We check these gradients against centered finite differences, which estimate the response from separate simulations with slightly increased and decreased parameter values.

For the three physical controls, we compare gradients of final mean atmospheric temperature over six integration lengths, from 1 to 148 steps (300--44{,}400\,s). We use three perturbation sizes spanning $10^{-3}$ to $10^{-5}$. All 54 comparisons pass the tolerance stated in Table~\ref{tab:derivative_checks}; the largest discrepancy at the longest horizon is $3.56\times10^{-7}$. The table also reports checks of the calibration and neural-training objectives. These comparisons verify derivative agreement at the selected states and integration horizons.

\begin{figure}[!hbt]
\centering
\includegraphics[width=0.85\linewidth]{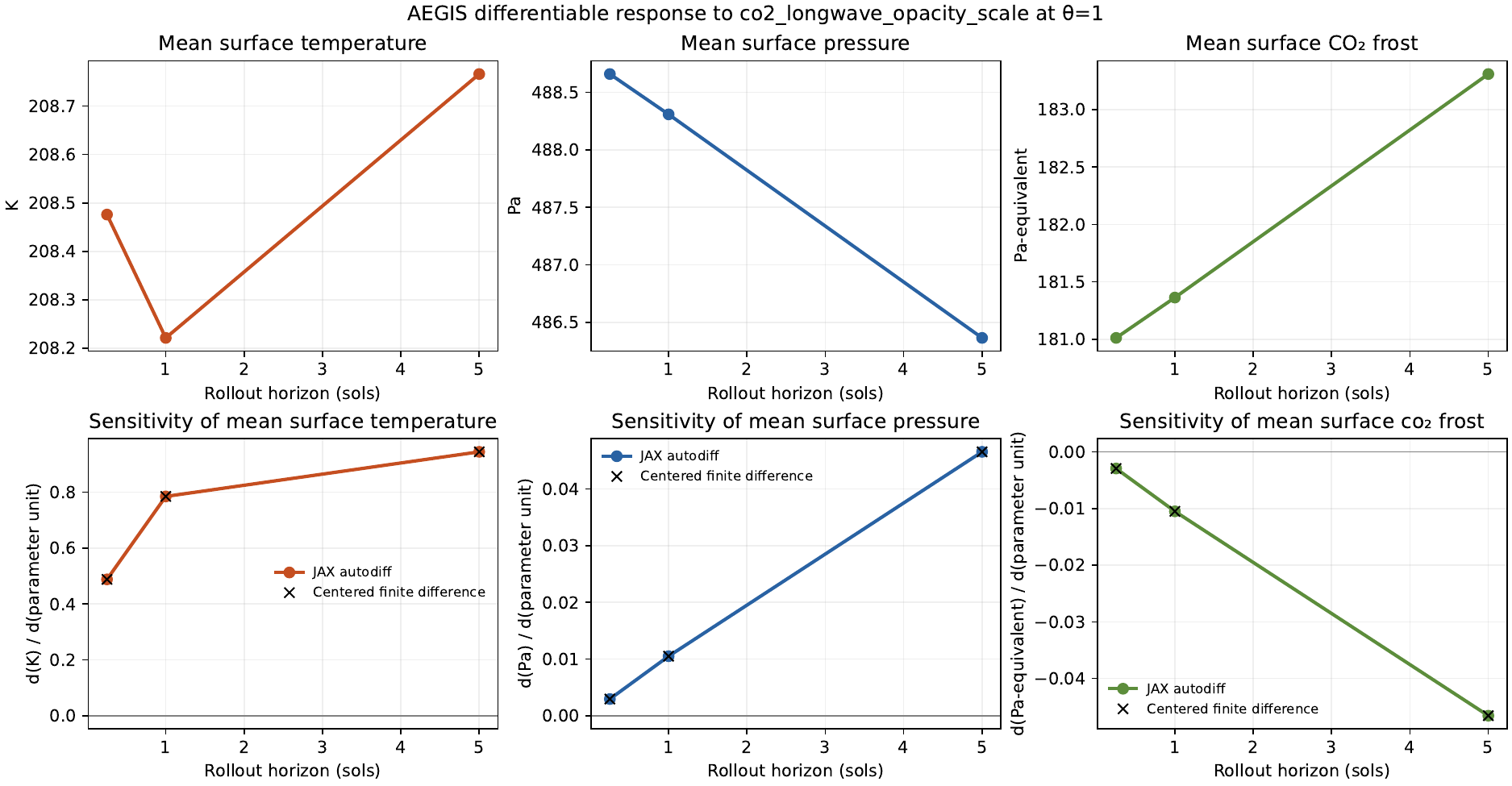}
\caption{Differentiable response of AEGIS to the \co{} longwave opacity multiplier $\beta$ at $\beta=1$, from a MY32, $L_s=45^\circ$ window. Top: global-mean surface temperature, surface pressure, and surface \co{} frost at rollout horizons of 0.25, 1, and 5 sols. Bottom: their derivatives with respect to $\beta$ computed by forward-mode automatic differentiation through the full coupled rollout (lines) and by centered finite differences (crosses). The pressure and frost sensitivities are equal and opposite, reflecting conservation of the atmosphere--frost \co{} inventory.}
\label{fig:co2-sensitivity}
\end{figure}

We also test how a change in \co{} longwave opacity affects surface temperature, pressure, and frost over several sols. The experiment starts from a Mars Year 32 (MY32) window at $L_s=45^\circ$ from ARCO-MACDA (a Mars atmospheric reanalysis archive) \citep{arcomacda,macda}. At the nominal opacity multiplier $\beta=1$, we evaluate all three sensitivities after 0.25, 1, and 5 sols (74, 296, and 1480 steps). All nine derivatives agree with centered finite differences using a perturbation of $10^{-3}$ - $10^{-5}$, with a maximum relative discrepancy of $3.99\times10^{-6}$ (Figure~\ref{fig:co2-sensitivity}).

The gradients describe the coupled response: increasing opacity warms the surface, suppresses condensation, and shifts \co{} from frost into the atmosphere. Surface-temperature sensitivity increases from 0.488 to 0.945\,K per unit $\beta$ between 0.25 and 5 sols, while pressure sensitivity increases from 0.00294 to 0.0466\,Pa per unit $\beta$. Pressure and pressure-equivalent frost sensitivities cancel to within $9.3\times10^{-11}$\,Pa. This checks the differentiated inventory constraint imposed by the projection. A warm evaluation of the three diagnostics and their derivatives over five sols takes about 92\,s on an NVIDIA A100 GPU.

We use trajectory gradients to calibrate three physical multipliers against reanalysis in twelve windows (Table~\ref{tab:skill}). Test temperature RMSE has a reduction of 7.84\%, with improvement in three of four test windows. Paired resampling of the held-out Mars Years 34--35 gives a 95\% percentile interval of 0.109--0.314\,K for the gain. The fitting and resampling protocols are given in Appendix~\ref{app:protocol}.

\input{assets/skill.tex}

\subsection{Computational performance and long integrations}
\label{sec:performance}

A cold-start integration over a full Mars year (668.6 sols) completes in 86.6 minutes on one NVIDIA A100 GPU, corresponding to about 463 simulated sols per wall-clock hour. The T21 configuration uses 12 sigma layers, a 450\,s time step, and float64 precision, requiring 131{,}900 steps. The timing includes compilation, checkpointing, and diagnostics. All 68 saved states remain finite, with a maximum relative atmosphere--frost \co{} inventory drift of $2.8\times10^{-9}$ under the per-step projection. Seasonal exchange lowers global-mean surface pressure to 485\,Pa at $L_s\approx146^\circ$ and raises it to 654\,Pa at $L_s\approx235^\circ$, a 30\% annual range. Surface frost reaches 185\,Pa in pressure-equivalent units.

A ten-Mars-year integration with the same configuration covers 6686 sols in 1{,}319{,}003 steps and completes in 11.4\,h on one NVIDIA A100 GPU. This corresponds to 586 simulated sols per hour, or about 68 minutes per Mars year. All 670 saved states remain finite, with a maximum relative \co{} inventory drift of $2.9\times10^{-8}$. Between years nine and ten, the reported differences in global-mean surface pressure and temperature are 0.16\,Pa and 0.20\,K, respectively.

The annual ablations test stability under changes to individual processes and numerical settings. All thirteen runs finish with finite saved states and inventory drift below $8.6\times10^{-9}$. They remove boundary-layer mixing, convection, regolith conduction, or \co{} exchange; weaken diffusion; halve the time step; or change each calibrated multiplier by $\pm20\%$.

To measure the size and effect of the inventory projection, we compare paired T21/L12 runs with atmospheric \co{} condensation and suspended-ice physics, using a 295.91748\,s time step. Here, the inventory includes atmospheric gas, surface frost, and suspended \co{} ice. Over the first year, the mean and maximum absolute pressure-rescaling corrections per step are $6.92\times10^{-8}$ and $1.94\times10^{-5}$ of the initial inventory. The cumulative absolute correction is 1.388\%, and the net correction is $-0.897\%$. With projection active, the maximum relative inventory drift is $4.00\times10^{-12}$. A paired run with numerical repair and rescaling disabled retains physical phase exchange and accumulates a positive inventory drift of 0.883\%. The two runs differ in global-mean surface pressure by 2.63\,Pa RMS, with a maximum absolute difference of 6.50\,Pa over the first year.

\subsection{Seasonal comparisons with MOLA topography}
The MOLA-based lower boundary produces a strong geographic structure in surface pressure \citep{smith2001, mola}. Elevated regions contain less atmospheric column mass, while low-lying terrain supports higher surface pressure. Table~\ref{tab:reference} reports the seasonal comparisons using a separately matched AEGIS average for each reference. High pressure correlations largely reflect the terrain structure. To assess agreement beyond this contribution, we fit log surface pressure against elevation separately for AEGIS and each reference, using area weights. Correlations between the remaining variations range from 0.794 to 0.900 across 16 season--year comparisons, compared with 0.978--0.990 for raw pressure.

\begin{figure}[!hbt]
\centering
\includegraphics[width=\linewidth,height=.58\textheight,keepaspectratio]{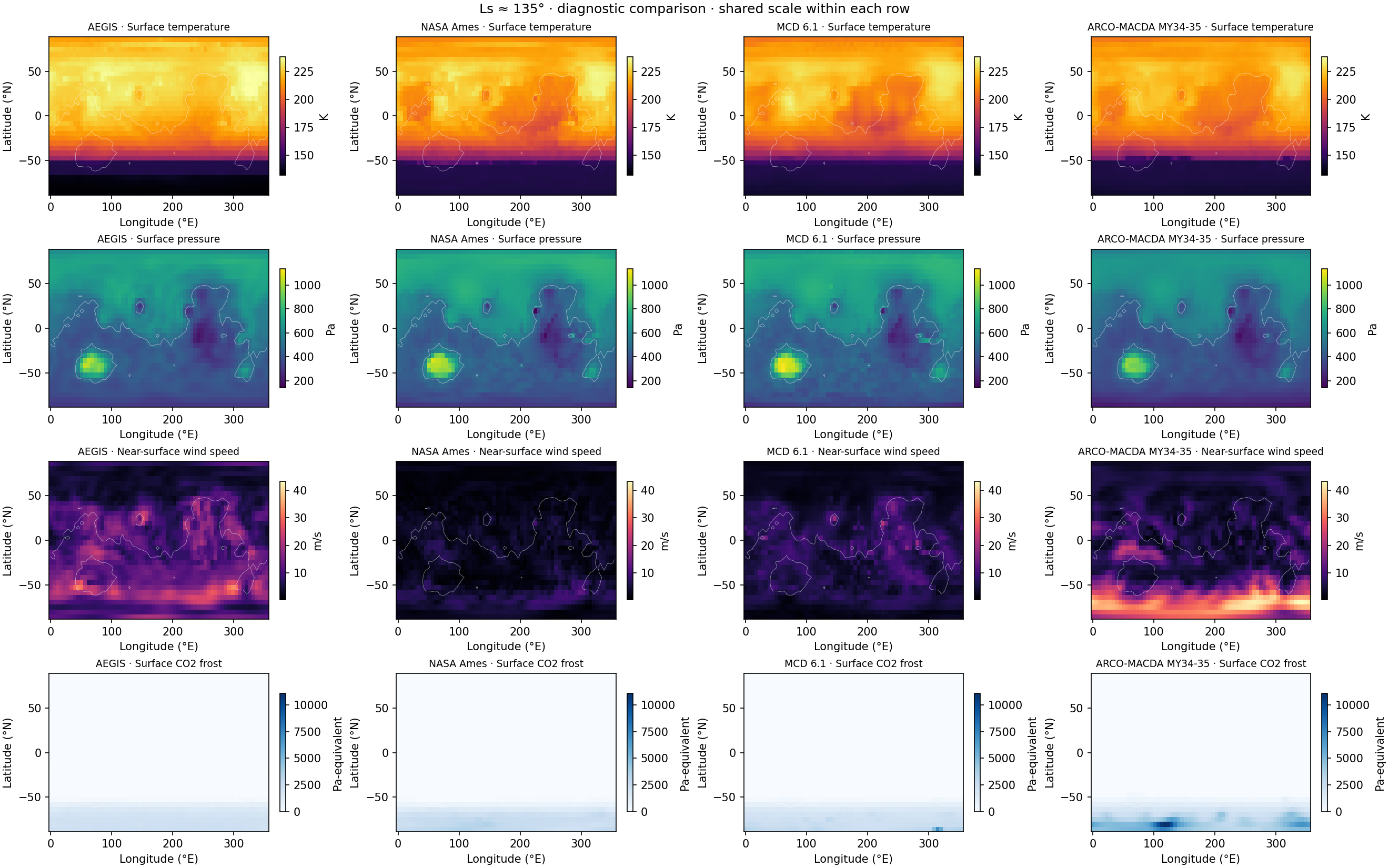}
\caption{Seasonal fields at $L_s=135^\circ$ from the AEGIS run and reference products. Columns show AEGIS, NASA Ames, MCD~6.1, and ARCO-MACDA; rows show surface temperature, surface pressure, near-surface wind speed, and surface \co{} frost in pressure-equivalent units. White contours mark zero MOLA elevation, and each row shares a color scale. AEGIS and Ames use matched five-sol intervals; MCD and ARCO-MACDA use the averaging described above.}
\label{fig:maps}
\end{figure}

Figure~\ref{fig:maps} shows fields from the second orbit of the two-Mars-year cold-start run. Frequent saved AEGIS fields are averaged over the same five-sol intervals as NASA Ames, with intervals weighted by duration. MCD averages four local solar times at each longitude. ARCO-MACDA uses daily means, averaged equally across MY34 and MY35. The single AEGIS column therefore matches Ames's temporal sampling; MCD and ARCO-MACDA retain different sampling. Wind heights also differ between products, and MCD frost is a monthly statistic. For the table, AEGIS is also sampled at the MCD local times and averaged over the ARCO-MACDA days; these separately matched fields differ from the single AEGIS column shown in the figure.

We compare the seasonal pressure swing with quality-screened daily observations from Viking Landers 1 and 2, using elevation-corrected model pressure and $5^\circ$ solar-longitude bins \citep{vikingDailyPressure}. AEGIS captures 98.5\% and 87.3\% of the observed peak-to-peak pressure swing at the two sites, respectively: 209.3 versus 212.5\,Pa at Viking 1 and 227.8 versus 260.8\,Pa at Viking 2. The modeled pressure maxima occur $25^\circ$ and $50^\circ$ earlier than observed. Prescribed dust differs from the Viking-year conditions.

\input{assets/reference.tex}

\subsection{Physics ablations: process sensitivity}
\begin{figure}[!hbt]
\centering
\includegraphics[width=0.9\linewidth]{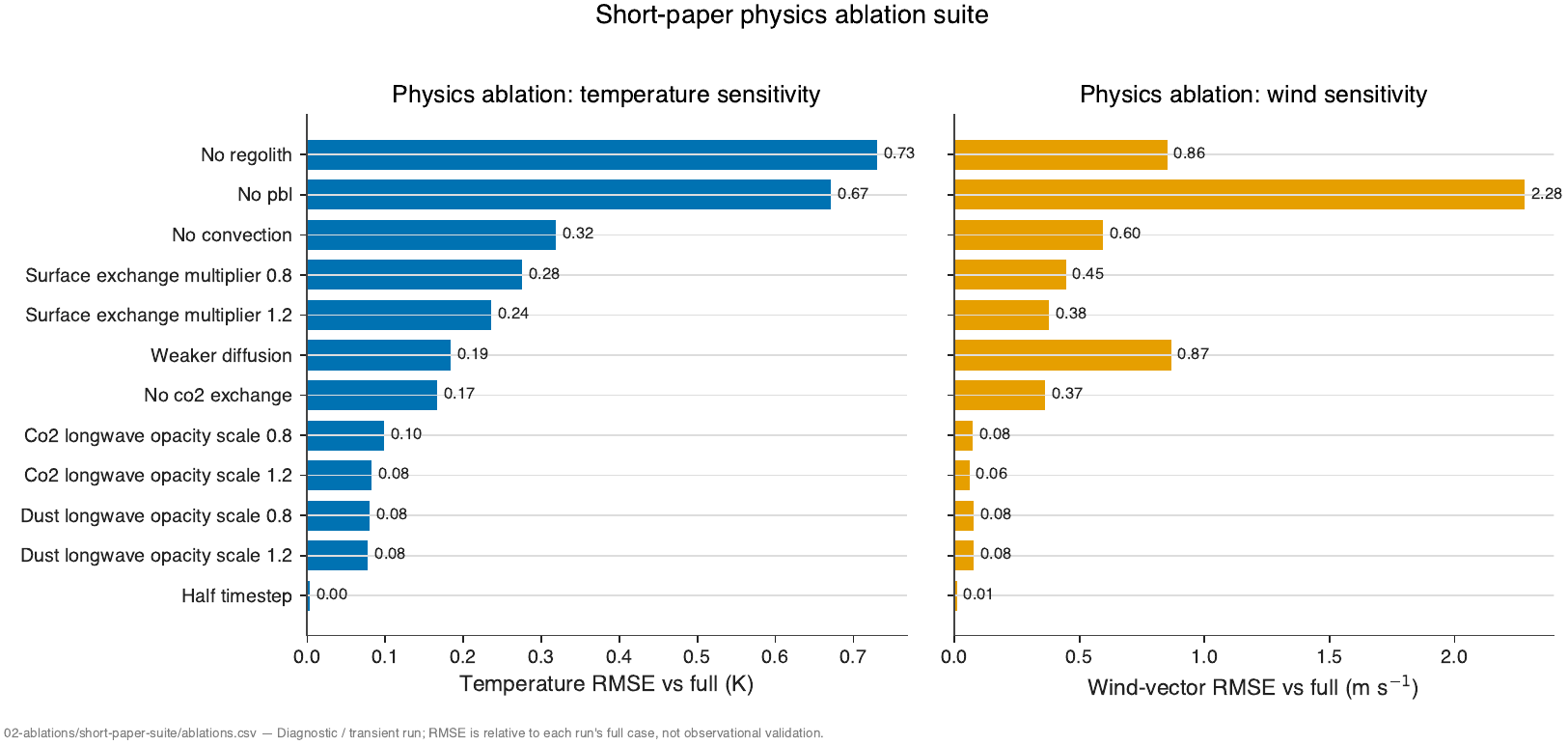}
\caption{Sensitivity of the T21/L12 Mars configuration. Bars show RMS differences from the paired full-physics trajectory over five saved times through 44\,400\,s; the right panel combines zonal and meridional wind RMSE by quadrature. All cases begin from the same state.}
\label{fig:short-physics}
\end{figure}

We experiment with an ablation suite that pairs the full Mars configuration with twelve process perturbations, such as removing boundary-layer mixing, dry convection, regolith conduction, or \co{} exchange, \co{} longwave opacity, and dust longwave opacity. Each is launched from the same initial state, so any divergence traces to the changed process, and we report the root-mean-square (RMS) separation from the paired full-physics trajectory. Across five saved times through 44\,400\,s, the perturbations show a sensitivity ranking (Fig.~\ref{fig:short-physics}): regolith conduction and boundary-layer mixing dominate, followed by dry convection and the surface-exchange multipliers, while the calibrated radiative opacity multipliers produce smaller but clearly resolved changes. Halving the time step from 300 to 150\,s gives an atmospheric-temperature RMS difference of 0.0042\,K across these five samples, smaller than every tested process perturbation.

All thirteen configurations complete with finite states, demonstrating stable long-horizon integration; by that horizon the freely evolving trajectories reach chaotic saturation, so RMS separations converge and process sensitivity is read from the short horizon. The restart-initialized full-physics ablation baseline uses a 300\,s time step and has a maximum relative atmosphere--frost \co{} inventory drift of $4.28\times10^{-9}$.

\subsection{Neural components as coupled experiments}
AEGIS provides the differentiable coupling and physical constraints needed to train neural components within a planetary model. The experiments illustrate how researchers can use these interfaces to replace a physical operator and fit physical parameters through simulated trajectories. Each example specifies the component, training objective, and evaluation protocol, providing a concrete starting point for studies using the framework.

\subsubsection{Learning radiative fluxes}
The radiation interface supports replacing a physical flux calculation with a trainable operator while preserving the discrete radiative budget. Researchers can fit the operator to local flux targets or train it through column trajectories. We exercise both paths using four-layer columns, starting from 100 local updates. The operator is then fine-tuned through 12-step trajectories for 20 updates or trained further on the local loss for the same wall-clock budget, allowing 234 updates. Table~\ref{tab:radiation_results} compares these configurations on interpolation and higher-dust extrapolation tests.

\begin{table}[!hbt]
\centering
\small
\caption{Final-state atmospheric-temperature RMSE (K) after 48 column steps for learned radiation variants. All models use the same structured flux interface.}
\label{tab:radiation_results}
\begin{tabular}{lcc}
\toprule
Training procedure & Interpolation & Higher dust \\
\midrule
Initial local fit & 0.620 & 1.799 \\
Trajectory fine-tuning & 0.566 & 1.653 \\
Continued local fitting & \textbf{0.416} & \textbf{0.979} \\
\bottomrule
\end{tabular}
\end{table}

\begin{figure}[!hbt]
\centering
\includegraphics[width=0.85\linewidth]{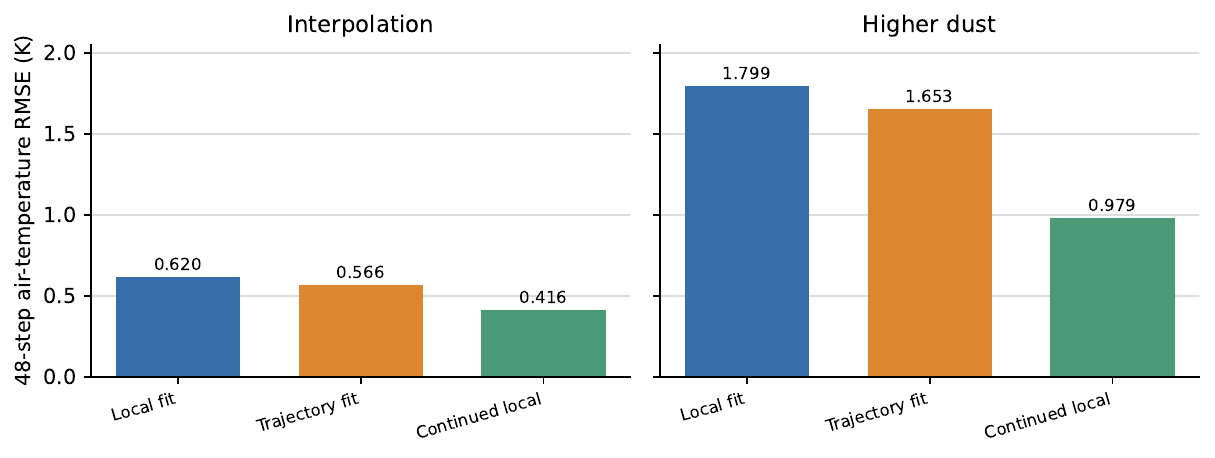}
\caption{Neural-radiation comparison after 48 column steps. Final-state air-temperature RMSE is evaluated against the compact physical radiation kernel. Interpolation and higher-dust evaluation each use 24 held-out columns. Bars compare the initial local fit, trajectory fine-tuning, and continued local fitting.}
\label{fig:neural-radiation}
\end{figure}

Continued local fitting gives the lowest 48-step temperature error under this budget (Table~\ref{tab:radiation_results}; Figure~\ref{fig:neural-radiation}). The largest instantaneous radiative-budget residual across held-out local evaluations is \(1.42\times10^{-14}\,\mathrm{W\,m^{-2}}\), and all three variants complete a two-step coupled T21 rollout. These checks exercise budget-preserving operator replacement, local and trajectory training, and short coupled execution.

\subsubsection{Recovering a surface heat-exchange parameter}
We test whether gradients through the column model can recover a known physical parameter from simulated temperatures. The parameter is a multiplier on the heat-transfer coefficient between the surface and the lowest atmospheric layer. The baseline heat-transfer coefficient is \(2\,\mathrm{W\,m^{-2}\,K^{-1}}\). We generate reference temperature trajectories with a multiplier of \(1.6\), then fit the multiplier from three starting values. Illumination, dust, and pressure remain fixed within each column.

\begin{figure}[!hbt]
\centering
\includegraphics[width=.75\linewidth]{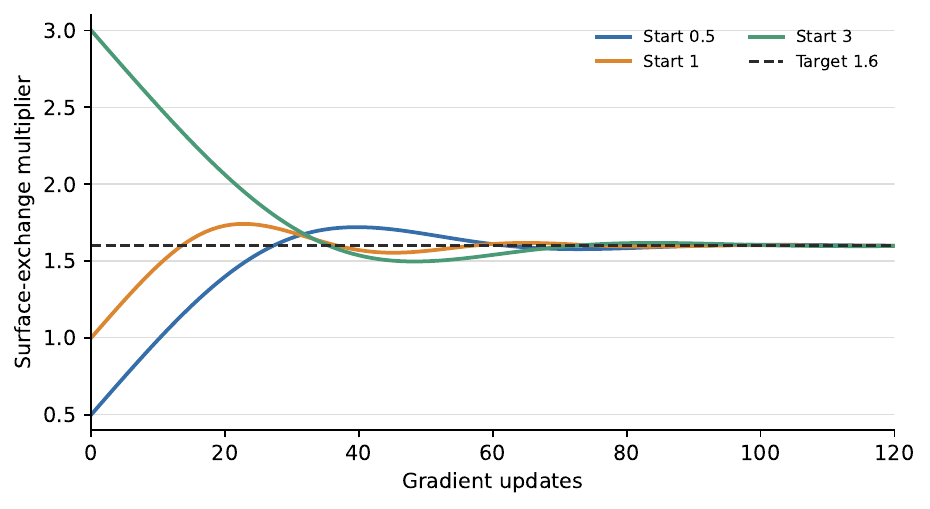}
\caption{Recovery of a surface heat-exchange multiplier in the column model. The prescribed value is 1.6. Three gradient-based fits start from 0.5, 1.0, and 3.0 and approach this value over 120 updates. Each curve shows the starting value and the estimate after every update.}
\label{fig:parameter-recovery}
\end{figure}

Each fit uses 120 gradient updates through 24 simulation steps. Starting from \(0.5\), \(1.0\), and \(3.0\), the fits recover \(1.59950\), \(1.59901\), and \(1.59696\), respectively (Figure~\ref{fig:parameter-recovery}). All three estimates are within \(0.20\%\) of the prescribed value. At multipliers \(0.5\), \(1.0\), and \(2.5\), the computed derivatives also agree with finite differences to relative discrepancies below \(3.41\times10^{-10}\). This experiment checks derivative accuracy and parameter recovery in a controlled setting. 

\section{Future extensions}

The generalization of this framework to other planetary atmospheres remains in scope. Future work will evaluate the framework on another well-documented atmosphere, such as Jupiter, using established benchmarks and reference data. Changing planetary constants alone does not validate the transfer of physical processes. Each planetary configuration would require appropriate thermodynamics, radiative properties, lower-boundary conditions, and condensable processes, followed by independent validation.

Further work on Mars includes broader observational comparisons, repeated neural experiments across seeds, and multi-year equilibration studies. These studies can reuse the coupled operators and trajectory objectives evaluated here.

Terraforming studies would extend the framework to changing Martian boundary conditions. Altered surface albedo, imposed heating, or specified changes in volatile inventory could be evaluated with spatial climate diagnostics. Differentiable objectives could connect these intervention parameters to regional temperature, frost persistence, or atmospheric pressure. Larger changes in atmospheric composition and mass would require corresponding changes to thermodynamics, opacity data, condensable processes, and the dynamical assumptions. These applications remain future work.

\section{Conclusion}                                                      
AEGIS couples Dinosaur's differentiable dynamics to Mars radiation, surface and regolith heat exchange, and seasonal \co{} mass transfer. Differentiation includes the atmosphere--frost inventory projection, while the learned radiation interface preserves the discrete column radiation budget.

The experiments verify gradients against finite differences and demonstrate physical calibration and neural training at the tested horizons. The ten-Mars-year run retains finite saved states, while reference comparisons identify remaining climate errors. Longer-term predictive validation, reliable coupled neural improvements, and applications to other planets or terraforming remain future work.

\subsubsection*{Acknowledgments}
This work was supported by the National Science Foundation under Grant No. 2414332

\bibliography{refs}
\bibliographystyle{iclr2027_conference}

\appendix
\section{Mars physics operators}\label{app:physics}

\subsection{Boundary data and radiative exchange}
MOLA elevation data define the lower boundary \citep{smith2001,mola}. The raster is regridded onto the transform grid and converted into spectral orography. Dust can be prescribed through seasonal or per-window fields. Boundary data are kept separate from the dynamical and physical operators.

The main Mars configuration uses Ames-derived 12-band correlated-$k$ tables, interpolated in temperature and log pressure, with two-stream transfer \citep{lacis1991,haberle2019ames}. The simplified column-learning experiments use the compact physical radiation kernel as their target generator and evaluate errors against its simulated temperatures.

Solar and thermal fluxes are evaluated at layer interfaces. Let $F_{k-1/2}$ and $F_{k+1/2}$ denote total net downward radiative flux at the top and bottom of layer $k$. Its radiative convergence and heat capacity per unit area are
\begin{equation}
 Q_k=F_{k-1/2}-F_{k+1/2},\qquad
 C_k=\frac{c_p p_s\Delta\sigma_k}{g},\qquad
 \left.\frac{dT_k}{dt}\right|_{\mathrm{rad}}=\frac{Q_k}{C_k}.
 \label{eq:app-radiative-heating}
\end{equation}
The surface receives the net flux at the bottom interface. Summing the atmospheric convergences and surface flux therefore gives the top-of-atmosphere net flux. The neural radiation interface uses this same accounting, so its discrete column radiation budget closes for any network weights. This identity accounts for radiative exchange. Full-model energy accounting also includes sensible, conductive, latent, and dynamical contributions.

\subsection{Surface exchange and the condensable reservoir}
Surface temperature responds to radiative, sensible, conductive, and latent heat exchange \citep{forget1999,haberle2019ames}. A bulk aerodynamic sensible heat flux couples the surface to the lowest atmospheric layer. The same flux is removed from one reservoir and added to the other. Surface stress transfers momentum; Richardson-number-dependent diffusion mixes momentum, heat, and tracers through a batched tridiagonal solve \citep{petrosyan2011pbl}. Kinetic energy dissipated by boundary-layer mixing is returned as heat. The separate hyperdiffusion filter provides numerical damping and can remove resolved kinetic energy. Dry convective adjustment conserves sigma-weighted enthalpy, and a 12-layer regolith supplies thermal memory.

Define $I=g\,m_{\mathrm{ice}}$, where $m_{\mathrm{ice}}$ is surface frost mass per unit area. Thus $I$ has units of pressure. Positive $dI/dt$ denotes condensation: frost increases and atmospheric pressure decreases. Sublimation reverses the signs. The local exchange terms satisfy
\begin{equation}
 \left.\frac{dp_s}{dt}\right|_{\mathrm{phase}}=-\frac{dI}{dt},
 \qquad
 \left.\frac{d\log p_s}{dt}\right|_{\mathrm{phase}}
 =-\frac{1}{p_s}\frac{dI}{dt}.
\end{equation}
The second expression converts the pressure tendency into the prognostic log-pressure tendency used by Dinosaur. A pressure-dependent saturation relation sets the frost temperature; the energy-limited exchange option also accounts for the available surface energy.

\subsection{Inventory projection and its derivative}
Finite Runge--Kutta updates of log pressure introduce inventory discretization error. After each complete step, negative frost is set to zero and the corresponding pressure-equivalent amount is removed from the atmospheric column. A global pressure rescaling then restores the prescribed total inventory.

For the closed surface-frost configuration, define the area-weighted mean $\langle q\rangle_A=\sum_i w_iq_i/\sum_i w_i$ and the pre-step inventory in pressure units as $J_n=\langle p_s^n+I^n\rangle_A$. After local frost repair, denote the fields by $\widetilde p_s$ and $I_+$. For feasible positive inventories, the global correction is
\begin{equation}
 p_s^{n+1}=c\,\widetilde p_s,\qquad
 c=\frac{J_n-\langle I_+\rangle_A}{\langle\widetilde p_s\rangle_A}.
 \label{eq:app-inventory-projection}
\end{equation}
Prescribed escape, when enabled, reduces the target inventory. The additional projection experiments in Section~\ref{sec:performance} include suspended \co{} ice in the total inventory and use a later physical configuration. Their correction budgets must be kept separate from the historical atmosphere--surface-frost drift values.

Automatic differentiation includes the repair and rescaling. In the surface-frost tests, pressure and pressure-equivalent frost sensitivities cancel because they differentiate the enforced constraint. This verifies the derivative of the enforced inventory constraint. Clipping and phase boundaries can change the active branch, so finite-difference agreement is assessed at the stated parameters and horizons.

\section{Calibration and neural-training protocol}\label{app:protocol}

\subsection{Data splits, initialization, and loss}
ARCO-MACDA supplies reanalysis targets for twelve prespecified windows \citep{arcomacda,macda}. MY denotes Mars year. Four windows from MY24, MY26, MY29, and MY31 form the training split; four from MY32--33 form validation; and four from MY34--35 form testing. Each split samples four seasonal quarters. Atmospheric temperature, winds, and pressure initialize the atmospheric state. Surface and frost use start-time fields; soil initialization uses preceding surface-temperature history. Dust and orbital phase are matched to the window.

Each fitting window spans the native reference interval of 7397.937\,s, integrated with 25 steps of 295.91748\,s, about one twelfth of a sol. Training minimizes squared errors in temperature and both horizontal wind components, scaled by training-target standard deviations. Horizontal Gaussian area weights, equal layer weights, and equal window weights are used. Reported split RMSE is the square root of mean per-window MSE. Training data determine normalization and fitted parameters, validation selects a checkpoint, and test data are reserved for evaluation.

\subsection{Physical calibration and uncertainty}
Bounded L-BFGS-B adjusts \co{} longwave opacity, dust longwave opacity, and surface exchange within $[0.05,2]$, $[0.05,2]$, and $[0.25,4]$, respectively. Three starting points require 56 objective-and-gradient evaluations in total. The recorded gradient and objective tolerances are $10^{-6}$ and $10^{-12}$. The training endpoints satisfy the projected-gradient criterion. Validation selects an earlier candidate with a training projected-gradient infinity norm of $2.02\times10^{-4}$.

The selected multipliers are approximately $(0.05,1.925,0.455)$. The \co{} opacity multiplier lies at its lower bound. These controls are fitted to the short-lead objective and can compensate for other model errors.

Test temperature RMSE experiences a fall of 7.84\%, with improvement in three of four windows. To describe uncertainty conditional on these windows, paired baseline and fitted errors are grouped by held-out Mars year. Resampling the two year blocks with replacement gives four ordered resamples. Recomputing the gain for each and taking percentile endpoints gives 0.109--0.314\,K. The interval describes variation among the two held-out year blocks for the selected fit. Some individual windows and wind components have higher errors after fitting.

\subsection{Coupled neural heating}
The heating pilot uses 50 Adam updates \citep{kingma2014}, learning rate $10^{-3}$, gradient-norm clipping at one, and physical multipliers fixed at one. A zero-output initialization reproduces its own physical baseline. The tanh output limits heating to $2\times10^{-4}$\,K\,s$^{-1}$. Table~\ref{tab:skill} retains separate baselines for physical calibration and neural heating because they belong to different experiments. The submitted heating pilot and its physical baseline both have held-out RMSE of 2.46\,K at the table's precision.

A separate three-seed follow-up has baseline RMSE 2.512949\,K and error reductions of $-0.0038$, $0.0104$, and $0.0281$\,K. The gains vary with seed. Evaluation uses two held-out year blocks, and training ends at the fixed update count. The reported seed variation describes the heating experiment.

\subsection{Radiation learning and its continuation controls}
The column experiments use four sigma layers, a 30\,s midpoint Runge--Kutta step, and a prescribed surface heat capacity of $6\times10^4$\,J\,m$^{-2}$\,K$^{-1}$. They evolve radiation and sensible exchange with fixed pressure, dust, and illumination per column. Circulation, regolith conduction, and \co{} phase change are disabled in this laboratory configuration. The tests assess radiation and sensible exchange in the reduced column model.

The historical radiation experiment uses 96 training columns and 24 columns in each validation and evaluation split. Whole columns belong to separate splits. Training and interpolation dust optical depths lie in $[0,0.8]$; the higher-dust test uses $[1,1.5]$. The network uses 17 input features, two 32-unit GELU hidden layers, and 20 raw outputs for four flux streams at five interfaces. Inputs include temperature and pressure profiles plus surface and forcing features. Physical boundary conditions are imposed before heating is computed.

An initial 100-update local fit is followed by either 20 trajectory updates through 12 steps or continued local fitting for the same measured training time. In this historical run, the latter performs 234 updates. Evaluation measures final-state atmospheric and surface temperature errors after 48 steps (1440\,s), beyond the 12-step training horizon (360\,s). Four states at steps 12, 24, 36, and 48 are saved for trajectory diagnostics and validity checks; the reported RMSE uses only the final state. The continuation methods receive equal measured training time and perform different numbers of updates. This budget depends on hardware. Their different objectives are a possible explanation for the ranking; identifying the cause requires a separate controlled comparison.

A separate follow-up uses three seeds and 20 updates for each continuation. It uses a later source revision and a fixed-update continuation budget. Its results correspond to that configuration:
\begin{table}[!hbt]
\centering\small
\caption{Follow-up column radiation errors after 48 steps. Entries are mean atmospheric-temperature RMSE $\pm$ sample standard deviation over three seeds, in K; each RMSE is measured at the final state after step 48. This fixed-update comparison is separate from the historical time-matched comparison in Table~\ref{tab:radiation_results}.}
\label{tab:app-radiation-seeds}
\begin{tabular}{lcc}
\toprule
Training & Interpolation & Higher dust \\
\midrule
Initial local fit & $0.839\pm0.188$ & $1.781\pm0.112$ \\
Trajectory continuation & $0.754\pm0.163$ & $1.704\pm0.056$ \\
Local continuation & $0.776\pm0.165$ & $1.652\pm0.130$ \\
\bottomrule
\end{tabular}
\end{table}
Trajectory continuation has the lower mean interpolation error, while local continuation has the lower mean higher-dust error. Seeds vary both generated data and initialization. The sample standard deviations describe their combined variation across the three runs.

The historical two-step T21 execution check is also distinct from a longer follow-up at T21/L4 with $3000\times30$\,s steps. In that test, all three variants pass the final-state validity checks for seed 0 and fail for seeds 1 and 2: six of nine deployments are flagged. Final-state air-temperature RMSE against the physical baseline is 18.5--19.3\,K for seed 0 and 64.3--90.2\,K for the others. The largest reported flux-budget residual is $2.27\times10^{-13}$\,W\,m$^{-2}$. The reports give flux closure, execution status, and final-state temperature errors for each deployment.

\subsection{Physical-parameter recovery and horizon limits}
The surface-exchange recovery example fits a multiplier on a 2\,W\,m$^{-2}$\,K$^{-1}$ bulk sensible conductance. This laboratory coefficient is distinct from the full GCM's aerodynamic closure. Reference trajectories use multiplier 1.6, and fitting uses 120 updates through 24 steps (720\,s). A bounded derivative-free scalar search reaches 1.6 in nine objective evaluations. This one-parameter experiment checks recovery through the differentiated column model alongside bounded scalar search.

The five-sol experiment in Section~\ref{sec:gradient-checks} holds the opacity multiplier at its nominal value and checks derivatives. Calibration instead uses the short reanalysis windows described above. The ten-year integration separately tests numerical behavior for its stated configuration. Assessing the long-term predictive accuracy of the fitted parameters requires held-out evaluation over longer trajectories and independent multiyear windows.

\section{Additional seasonal reference comparisons}\label{app:seasonal-maps}
\begin{figure}[!hbt]
\centering
\includegraphics[width=\linewidth,height=.58\textheight,keepaspectratio]{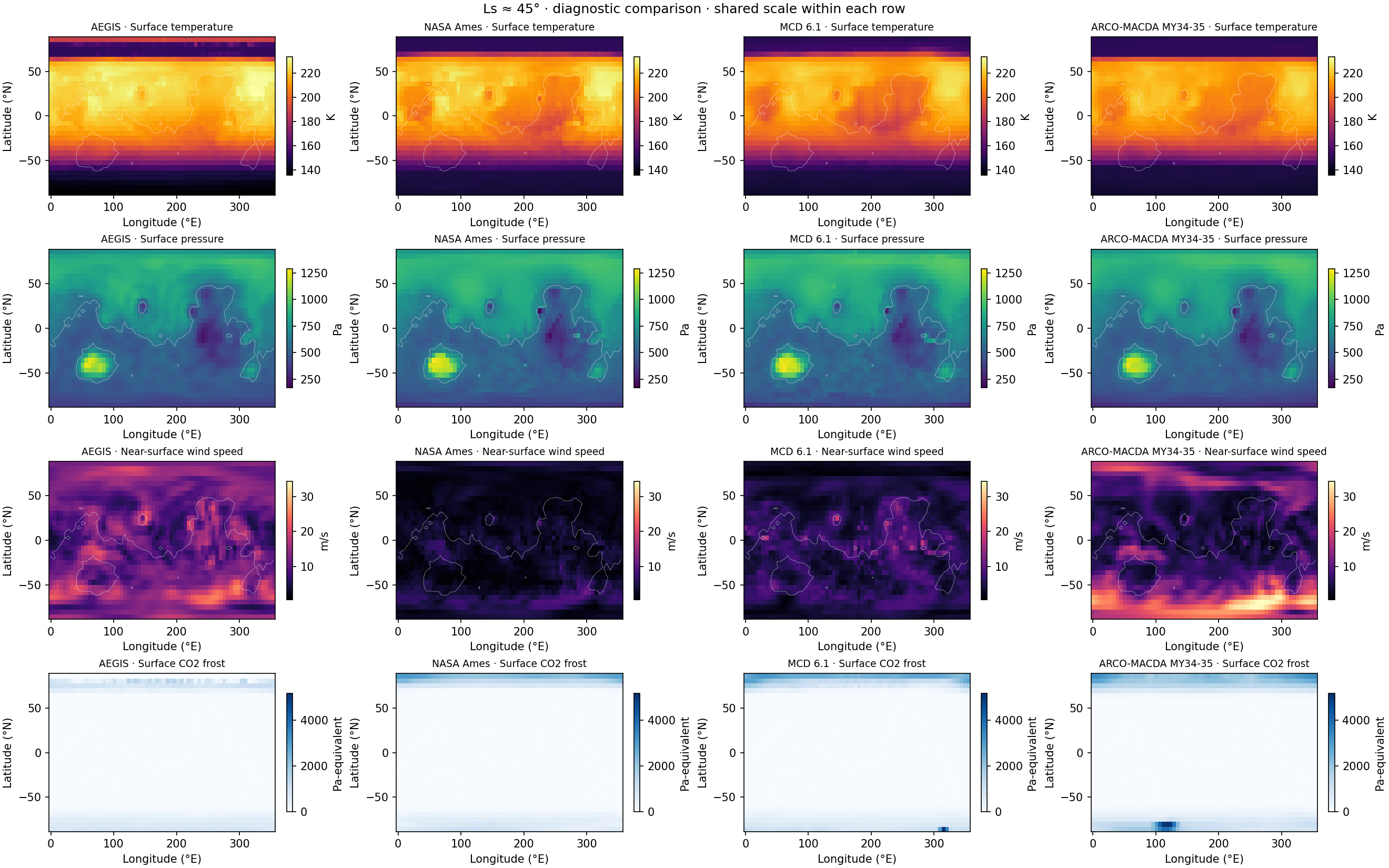}
\caption{Seasonal fields at $L_s=45^\circ$. Columns, rows, MOLA contours, and averaging follow Figure~\ref{fig:maps}.}
\label{fig:maps-ls045}
\end{figure}

\begin{figure}[!hbt]
\centering
\includegraphics[width=\linewidth,height=.58\textheight,keepaspectratio]{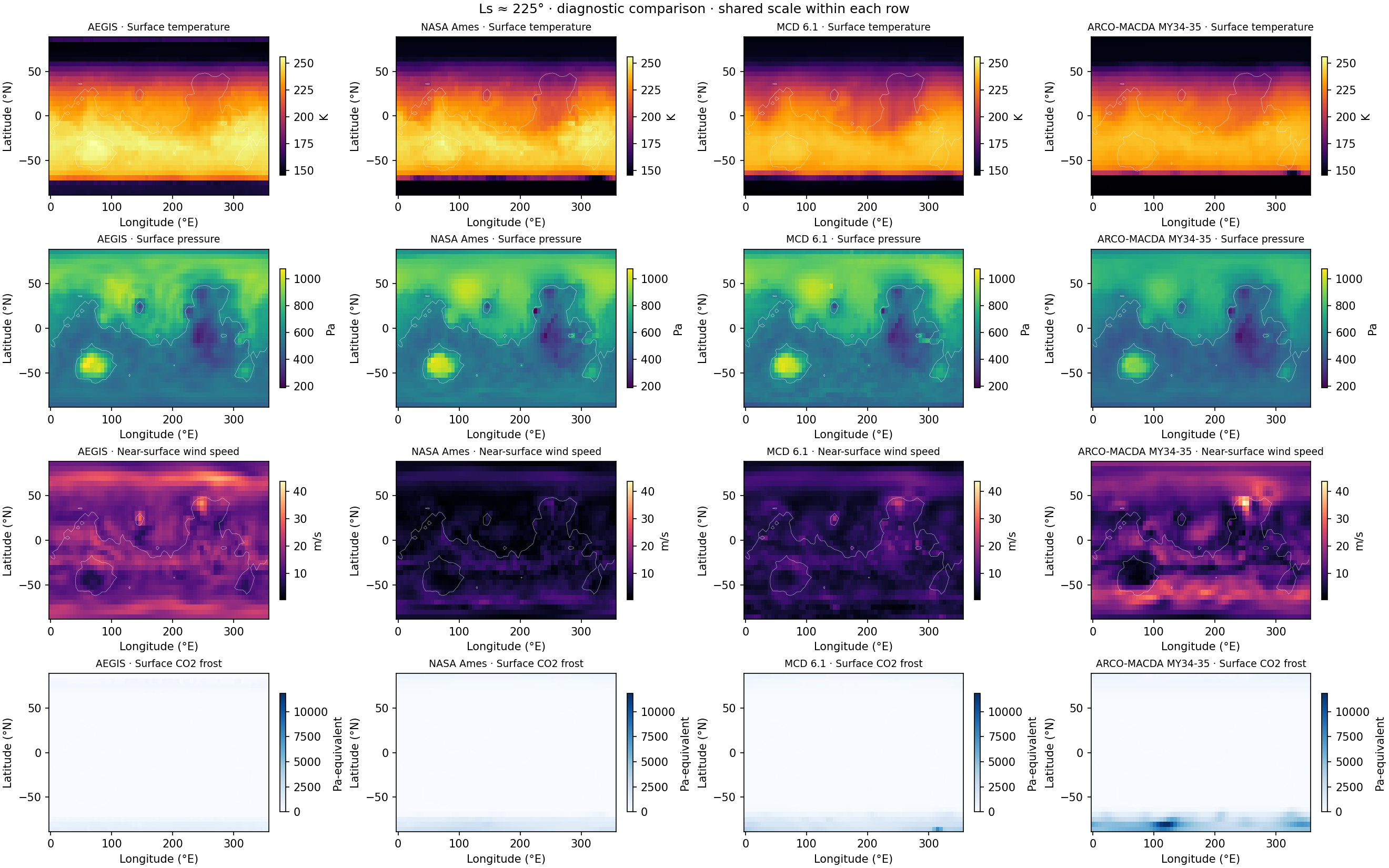}
\caption{Seasonal fields at $L_s=225^\circ$. Columns, rows, MOLA contours, and averaging follow Figure~\ref{fig:maps}.}
\label{fig:maps-ls225}
\end{figure}

\begin{figure}[!hbt]
\centering
\includegraphics[width=\linewidth,height=.58\textheight,keepaspectratio]{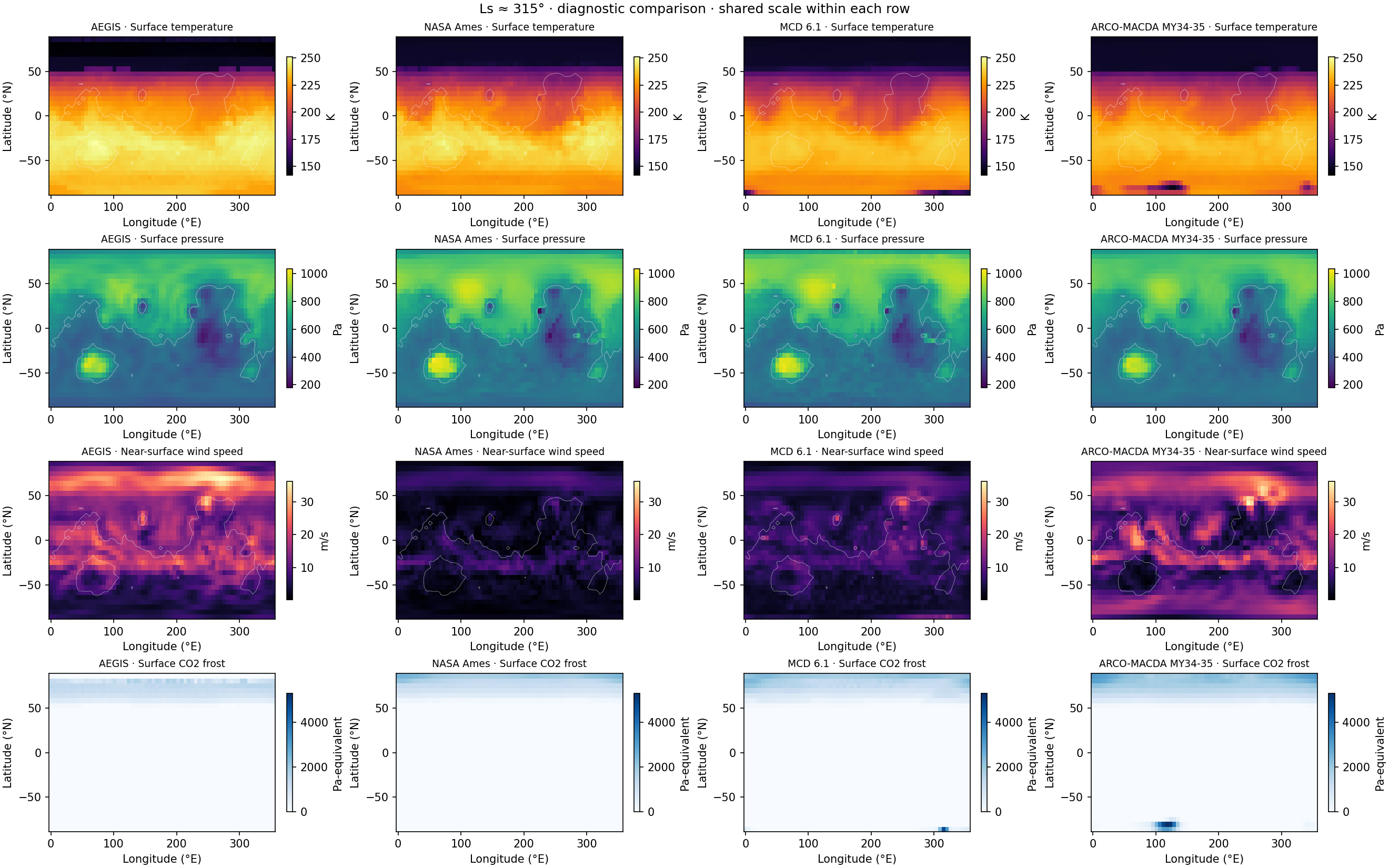}
\caption{Seasonal fields at $L_s=315^\circ$. Columns, rows, MOLA contours, and averaging follow Figure~\ref{fig:maps}.}
\label{fig:maps-ls315}
\end{figure}

Figures~\ref{fig:maps-ls045}, \ref{fig:maps-ls225}, and~\ref{fig:maps-ls315} supply the other three seasons alongside $L_s=135^\circ$ in Figure~\ref{fig:maps}, giving four seasons overall. Each row shares a color scale within a figure; limits can differ between seasons.

\subsection{Temporal sampling and spatial metrics}
Each reference is compared with its own matched AEGIS average. Approximately hourly model fields are integrated over the Ames five-sol and ARCO-MACDA daily intervals using trapezoidal quadrature and interpolated boundaries. MCD uses four local solar times sampled separately at each longitude. Table~\ref{tab:reference} averages four seasonal metric values; for ARCO-MACDA, MY34 and MY35 maps are averaged before computing those seasonal metrics. The single AEGIS image column uses the Ames averaging; the table uses reference-specific AEGIS averages. Wind heights and the monthly MCD frost statistic also differ.

For pressure, log pressure is independently regressed on elevation for the model and each reference with area weights. Correlating the residuals gives 0.594--0.800, compared with 0.978--0.990 for raw pressure. These statistics describe the residual pressure variations after removing the fitted elevation trend.

\subsection{Viking station comparison and practical scope}
Model pressure is interpolated to each Viking station and corrected from model-grid elevation to local MOLA elevation using a hydrostatic relation and lowest-layer temperature. Daily model means are matched to observed solar longitudes, then averaged with quality-screened PDS daily observations in $5^\circ$ bins \citep{vikingDailyPressure}. Seasonal anomaly correlations are 0.959 and 0.950, while pressure RMSE is 56.8 and 76.0\,Pa at Viking 1 and 2. The early pressure maxima reported in the main text remain a discrepancy. The comparison uses prescribed model dust and Viking mission observations. The reported pressures describe the two station locations.

These T21/L12 comparisons assess coarse global and seasonal structure. Saved-output frequency sets temporal sampling; the transform grid sets spatial resolution. Landing-site forecasts and operational decisions require their own resolution, error measures, and validation. The reference fields come from a cold-start integration. Multiyear equilibration and applications beyond Mars remain future work.

\section{Numerical settings, horizons, and reproducibility}\label{app:numerics}

\subsection{Spectral discretization, diffusion, and time stepping}
T21/L12 denotes spherical-harmonic horizontal dynamics with 12 sigma layers. The $64\times32$ longitude--latitude array is a transform grid for nonlinear terms and column physics. Parameter and neural-direction finite differences independently check gradients of this spectral model.

SIL3 treats the fast linear gravity-wave terms implicitly. The post-step eighth-order filter for a modal variable is
\begin{equation}
 \widehat q_{\ell m}^{\,n+1}
 =
 \exp\!\left[-\frac{\Delta t}{\tau}
 \left(\frac{|\lambda_\ell|}{\Lambda_{\max}}\right)^4\right]
 \widehat q_{\ell m}^{\,*},
 \qquad \Lambda_{\max}=\max_j|\lambda_j|,
 \label{eq:app-diffusion}
\end{equation}
where $\lambda_\ell$ is a spherical-Laplacian eigenvalue and the star denotes the post-integrator state. The fourth power corresponds to eighth-order hyperdiffusion. The damping time is $\tau=0.1$ sol, or 8877.5244\,s; weaker diffusion uses 0.2 sol. The filter acts on vorticity, divergence, and temperature variation, excluding pressure, tracer mixing ratios, and surface reservoirs.

A sampled horizontal advective Courant diagnostic is
\begin{equation}
 C_h=\max_{\mathrm{grid,\ layers}}\Delta t
 \left(\frac{|u|}{a\cos\phi\,\Delta\lambda_{\min}}
       +\frac{|v|}{a\,\Delta\phi_{\min}}\right),
 \label{eq:app-cfl}
\end{equation}
where $a$ is radius, $\phi$ is latitude, and the angular spacings come from the transform grid. Separate annual tests of the updated physical configuration at 450, 300, and 150\,s give sampled maxima of 1.90, 1.39, and 0.677. All complete with finite diagnostics. Their time-integrated field-temperature RMS separations are 3.466\,K (450/300\,s), 3.309\,K (450/150\,s), and 3.272\,K (300/150\,s). Peak local Mach numbers are approximately 1.07--1.11. These diagnostics describe sampled horizontal advection and trajectory separation. Full three-dimensional stability and climatological convergence require further assessment. Advection and explicit physics constrain the time step.

The historical short ablation comparison is different: its 0.0042\,K step-halving result pools temperature errors over 6000, 12000, 22200, 30000, and 44400\,s, with Gaussian horizontal weights and equal layer and sample weights. It compares 300 with 150\,s. The separate annual ablation has a single-sample difference that also rounds to 0.0042\,K at 88800\,s, but its 74-sample annual RMS is 3.700\,K. The short-horizon, endpoint, and annual statistics use distinct averaging intervals.

\subsection{Distinct inventory diagnostics}
The two historical annual drifts refer to different experiments. The $2.8\times10^{-9}$ maximum relative drift belongs to the 450\,s cold-start timing run with 68 saved states, initialized at rest with uniform 200\,K atmosphere, surface, and soil and no frost. The $4.28\times10^{-9}$ value belongs to the 300\,s restart-initialized ablation baseline with 74 samples and physical multipliers $(0.6,0.8,1.4)$. Both use $|M(t)/M(0)-1|$ at saved diagnostic times. Each value belongs to its stated initialization, time step, and diagnostic sampling. The later projection-on/off comparison, which includes suspended ice, supplies a third, separately labelled inventory diagnostic.

\subsection{Hardware and interpretation of cost}
Reproduction uses each experiment's corresponding configuration, checkpoints, and source records.
\begin{table}[!hbt]
\centering\small
\caption{Confirmed hardware for the experiment groups. Reported memory values give device capacity.}
\label{tab:app-hardware}
\begin{tabular}{p{0.40\linewidth}p{0.31\linewidth}p{0.14\linewidth}}
\toprule
Experiment group & Recorded device & Precision \\
\midrule
Annual and ten-year timings and five-sol sensitivity & NVIDIA A100 & float64 \\
Verified short-lead calibration and later paired projection runs & NVIDIA A100, 40\,GB & float64 \\
\bottomrule
\end{tabular}
\end{table}
The column reports record Python 3.12.3 and JAX 0.11.0. AEGIS timings include compilation, checkpointing, and diagnostics as stated in Section~\ref{sec:performance}. The later paired projected run records about 823 simulated sols per hour over 1338 sols on an A100 PCIe 40\,GB; its integration-loop timing includes compilation and sampled output and excludes staging and setup. It is a different configuration from the historical timing runs.

Forward throughput and derivative cost describe separate workloads. Calibration evaluates physical-parameter derivatives with forward mode; neural training uses reverse mode. Rematerialization reduces reverse-mode storage by recomputing forward intermediates. Time-step restrictions follow the numerical scheme, and derivative accuracy is assessed at the stated horizons. Throughput is reported for AEGIS with its hardware, resolution, time step, physics, and timed region.

\end{document}

%% file: math_commands.tex
\usepackage{amsmath,amsfonts,bm}

\def\eqref#1{equation~\ref{#1}}

\def\1{\bm{1}}

\def\vx{{\bm{x}}}
\def\vy{{\bm{y}}}
\def\vz{{\bm{z}}}

\DeclareMathAlphabet{\mathsfit}{\encodingdefault}{\sfdefault}{m}{sl}
\SetMathAlphabet{\mathsfit}{bold}{\encodingdefault}{\sfdefault}{bx}{n}

%% file: assets/skill.tex
\begin{table}[t]
\centering\small
\caption{Atmospheric temperature RMSE (K) at a lead of 7397.937\,s. Each split contains four seasonal windows, combined as the square root of their mean squared errors. Each fitting experiment is paired with its own physical baseline.}
\label{tab:skill}
\begin{tabular}{lrrr}
\toprule
Configuration & Training & Validation & Test \\
\midrule
\multicolumn{4}{l}{Physical calibration} \\
Physical baseline & 2.48 & 2.56 & 2.53 \\
Calibrated physics & 2.27 & 2.38 & 2.33 \\
\midrule
\multicolumn{4}{l}{Coupled neural heating} \\
Physical baseline & 2.45 & 2.49 & 2.46 \\
Neural heating & 2.42 & 2.49 & 2.46 \\
\bottomrule
\end{tabular}
\end{table}

%% file: assets/reference.tex
\begin{table}[t]
\centering\small
\caption{Seasonal reference comparisons. Entries are arithmetic means of four seasonal RMSE values and pattern correlations for surface pressure and temperature. Each reference uses its own temporally matched AEGIS average. For ARCO-MACDA, metrics are computed after averaging the matched MY34 and MY35 maps. High pressure correlations largely reflect terrain structure.}
\label{tab:reference}
\begin{tabular}{lrrrr}
\toprule
& \multicolumn{2}{c}{Pressure} & \multicolumn{2}{c}{Temperature} \\
Reference & RMSE (Pa) & Correlation & RMSE (K) & Correlation \\
\midrule
NASA Ames & 41.89 & 0.9873 & 8.01 & 0.9795 \\
MCD 6.1 & 46.88 & 0.9793 & 11.20 & 0.9725 \\
ARCO-MACDA & 42.43 & 0.9888 & 11.11 & 0.9777 \\
\bottomrule
\label{mola-table}
\end{tabular}
\end{table}